\documentclass[aps,prl,twocolumn,amsmath,amssymb]{revtex4-2}
\usepackage{graphicx}
\usepackage{dcolumn}
\usepackage{bm}
\usepackage{amsmath}
\usepackage{amsthm}
\usepackage{amssymb}
\usepackage{amsfonts}

\usepackage{soul}
\usepackage{color}

\usepackage{graphicx}
\usepackage{multirow}
\usepackage{stackrel}
\usepackage{mathtools}
\usepackage{wrapfig}
\usepackage{pgfplots}
\usepackage{subfig}
\pgfplotsset{compat=1.9}

\begin{document}
\title{
Anomalous localization of light in one-dimensional L\'evy photonic lattices}
\author{Alejandro Ram\'irez-Yañez$^1$, Thomas Gorin$^{1,2}$, Rodrigo A. Vicencio$^{3,4}$, V\'ictor A. Gopar$^{5}$}
\affiliation{$^1$Departamento de F\'isica, CUCEI, Universidad de Guadalajara, Guadalajara, Jalisco, C.P. 44430, México.}
\affiliation{$^2$Max-Planck-Institut für Physik komplexer Systeme, Nöthnitzer Str. 38, D-01187 Dresden, Germany\looseness=-1}
\affiliation{$^3$Departamento de F\'isica, Facultad de Ciencias F\'isicas y Matem\'aticas, Universidad de Chile, Santiago 8370448, Chile}
\affiliation{$^4$Millenium Institute for Research in Optics–MIRO, Universidad de Chile, Chile.}
\affiliation{$^5$Departamento de F\'isica Te\'orica and BIFI, Universidad de Zaragoza, Pedro Cerbuna 12, E-50009, Zaragoza.}


\begin{abstract}
Localization of coherent propagating waves has been extensively studied over the years, primarily in homogeneous random media. However, significantly less attention has been given to wave localization in inhomogeneous systems, where the standard picture of Anderson localization does not apply, as we demonstrate here. We fabricate photonic lattices with inhomogeneous disorder, modeled by heavy-tailed $\alpha$-stable distributions, and measure the output light intensity profiles. We demonstrate that the spatial localization of light is described by a stretched exponential function, with a stretching parameter $\alpha$, and an asymmetric localized profile with respect to the excitation site. We support our experimental and theoretical findings with extensive tight-binding simulations.
\end{abstract}
\maketitle
Wave interference has been recognized as a fundamental phenomenon in physics since the pioneering works by Young and Fresnel in the 19th century when the classical wave theory of light was under development~\cite{Kipnis_1994}. Later, in the late 1950s, Anderson proposed an unexpected consequence of quantum wave interference: non-interacting electrons in random media could be spatially localized due to destructive interference of the electron wave functions~\cite{Anderson_1958}.  
The phenomenon of localization is now understood as a general phenomenon of coherent wave interference, which has been observed in several scattering setups involving classical electromagnetic, acoustic, flexural, and quantum waves~\cite{Cutler_1969,Dalichaouch_1991,Berry1997,Chabanov_2000, Schwartz_2007,Lahini_2008,Hu_2008,Flores_2013}. For a review, see, for instance~\cite{Lee_1985,Kramer_1993,Abrahams_2010}. 

Most studies on Anderson localization assume homogeneous disorder, where the asymptotic behavior of the wave envelope decays exponentially in space.   
It is of general knowledge that in one dimension (1D), any amount of disorder localizes waves that are initially spreading in a medium~\cite{Lee_1985, Kramer_1993, Abrahams_2010}; i.e., wave diffusion eventually ceases if the system is sufficiently large. In 1D, all eigenstates are strongly localized, and the exponential decay of the wave function is governed by the localization length, $\xi$, as $|\psi|^2 \sim \exp{\left( -|x-x_0|/\xi\right)}$, for large $|x-x_0|$. Although the assumption of homogeneous disorder in random media is a convenient and useful simplification, inhomogeneous disorder is common in both natural and synthetic materials. Furthermore, although the presence of the disorder is typically seen as a disadvantage, controlling it in random media can be leveraged to enhance image transport or cloaking effects~\cite{Mosk2012, Schittny2016}, for instance.
Thus, investigating the diffusion and localization of waves in inhomogeneous random media is of fundamental and practical interest. 
\begin{figure}
    \begin{center}
    \includegraphics[width=1.\columnwidth]{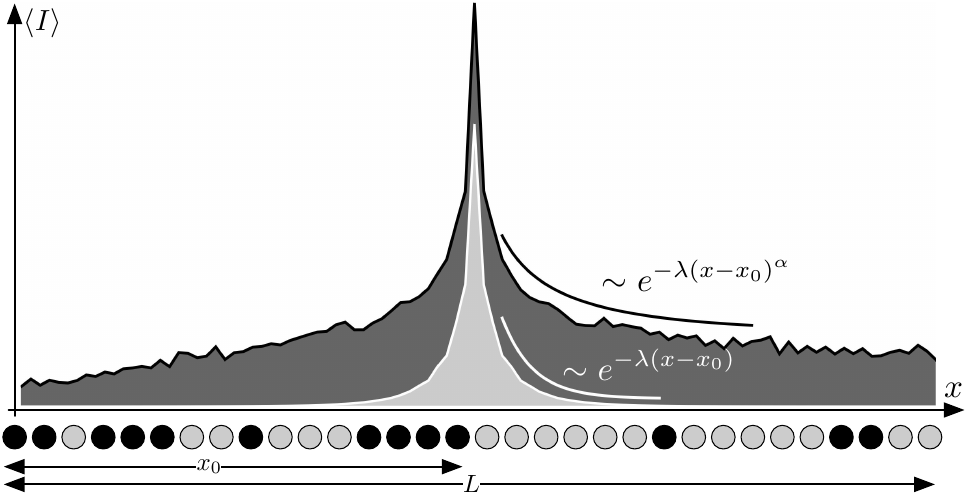}
    \caption{Sketch of the averaged intensity profile for homogeneous (light-gray) and inhomogeneous (dark-gray) disorder distributions, in a system of length $L$. The sequence of black disks represents scatterers of a particular realization of inhomogeneous disorder. For homogeneous disorder (Anderson case), light is exponentially localized, whereas for inhomogeneous disorder, with $\alpha$-stable distributed scatterers, a stretched exponential with exponent $\alpha$ is observed.}
    \label{fig1}
    \end{center}
\end{figure}

Photonic lattices have emerged as a key experimental tool for exploring and demonstrating fundamental phenomena related to the propagation of waves in periodic and aperiodic finite systems. Anderson localization has been observed in photorefractive-induced 2D disordered lattices~\cite{Schwartz_2007}, showing an exponential decay in the intensity profile. The optical problem can be directly mapped to a one-particle problem in quantum mechanics, allowing the wave function to be directly imaged using CCD cameras~\cite{Lederer_2008}. Studies on lattices that include disorder and nonlinearity have shown enhanced localization~\cite{Schwartz_2007,Lahini_2008,Naether_2013}.
Although some years ago there was intense activity in the study of disordered systems, nowadays, research on localization phenomena is mainly focused on fully periodic lattices with geometrical~\cite{FlachFB,Vicencio_2021} or topological~\cite{Ozawa_2019} properties, among others.

In this work, we study theoretically and experimentally the diffusion of waves in inhomogeneous random media. We measure output intensity profiles in several fabricated photonic lattices and investigate the effects of inhomogeneous disorder on the localization of light. A family of heavy-tailed distributions known as $\alpha$-stable or L\'evy  stable distributions~\cite{Uchaikin_2011} is used to model the inhomogeneous disorder in the fabricated lattices. We demonstrate that light is localized according to a stretched exponential function, which differs from the purely exponential localization observed in the standard Anderson regime. This effect is illustrated in Fig.~\ref{fig1}, where we observe the differences in the averaged decaying profiles. We also remark that the localization is not symmetric with respect to $x_0$, the site where the wave packet is released.

The heavy tails of the L\'evy stable distributions give a significant probability that light undergoes what is known as L\'evy random walks. This means that light can travel long distances without scattering, which has a crucial impact on intensity statistics,  as we show below. L\'evy random processes appear in various contexts. For instance, the L\'evy  distributions have been used to describe the large random fluctuations of the stock market prices \cite{Mandelbrot_1963}, while evidence of foraging large random displacements of different marine predators and human hunter activities can be described by L\'evy $\alpha$-stable distributions~\cite{Raichlen_2013, Sims_2008}. The effects of disorder modeled by heavy-tailed distributions on the transmission of classical and quantum waves have been studied theoretically and experimentally~\cite{Leadbeater_1998,Boose_2007, Barthelemy_2008, Beenakker_2009,  Mercadier_2009, Burioni_2010,  Falceto_2010,Eisfeld_2010, Sibatov_2011, Amanatidis_2012, Vlaming_2013, Fern_ndez_Mar_n_2012,Fern_ndez_Mar_n_2014,Asatryan_2018,Lima_2019,Razo_L_pez_2020}. 

An $\alpha$-stable distribution $\rho_\alpha(z)$ is essentially characterized by its power-law tail; i.e., 
$\rho_\alpha(z) \sim 1/|z|^{1+\alpha}$ for $|z| \gg 1$ where $0 < \alpha < 2$. 
Due to the slow decay of the tails, the first and second moments of the L\'evy distribution diverge for $\alpha < 1$. We will focus on this case, where the effects of the heavy tails are more pronounced. Also, the slow power-law decay is crucial for observing rare random events with non-negligible probability. 
These rare events can be so important that they could determine the complete statistics of a given random process. 

We consider the scattering problem of a 1D structure composed of randomly separated scatterers to show the unconventional localization of waves for inhomogeneous disorder. Thus, waves are injected at a position $x_0$ into the sample, while the intensity $I(x)$ is obtained as a function of position $x$, as described in Fig.~\ref{fig1} (for more details, see the Supplemental Material~\cite{SM}). Assuming a small empty region around $x$, the intensity is determined by the modulus square of the sum of the left and the right traveling waves as
\begin{equation}
\label{I}
 I(x)=|c_{+} \exp{(ikx)}+ c_{-} \exp{(-ikx)}|^2\ , 
\end{equation}
where $k$ is the wavenumber, and $c_{+}$ and $c_{-}$  are the wave amplitudes (the common factor $\sqrt{2\pi \hbar^2 k/m}$ of the amplitude waves is not included). 
A source of disorder is introduced by considering a random position of the scatterers. 
The statistical properties of $I(x)$ are thus calculated by averaging different realizations of the disorder. The theoretical description of the wave statistics inside inhomogeneous 1D disordered media is based on random matrix and localization scaling theory~\cite{Anderson_1980, Mello_2004, Beenakker_1997}. The details of the calculations are provided in~\cite{SM}. We study the average of the logarithmic intensity since it allows us to derive simple analytical expressions in terms of physical quantities, such as the transmission, and it reveals the intensity's distinctive behavior along 1D inhomogeneous disordered media.

We assume that random distances between scatterers in standard homogeneous media follow a light-tailed distribution, for example, a Gaussian distribution ($\alpha=2$). For homogeneous random media,  we expect to observe the well-known exponential localization of waves, as Anderson predicted for electrons~\cite{Anderson_1958}.  Indeed,  
a linear dependence of the logarithmic intensity with the observation's position $x$ is obtained: 
\begin{equation}
\label{lnI_Anderson} 
 \langle \ln I(x) \rangle =-|x-x_0|/\ell\ , 
\end{equation}
where $\ell$ is the mean free path. For $x_0=0$, Eq.~(\ref{lnI_Anderson}) reduces to the result obtained in~\cite{Cheng_2017} for incident waves from the left side of the sample. The linear behavior of $ \langle \ln I(x) \rangle$ with the position $x$ is a signature of Anderson localization. Thus, it is expected an exponential decay of the intensity for large distances $\left|x-x_0 \right|$:
\begin{equation}
\label{I_Anderson}
 \langle I(x) \rangle \sim e^{- \left|x-x_0 \right|/\xi}\ .
\end{equation}
Since in 1D systems with homogeneous disorder $\xi =2 \ell$~\cite{Thouless_1973, Scott_1985}, from Eqs.~(\ref{lnI_Anderson}) and (\ref{I_Anderson}), we note that the typical intensity $\exp{\langle \ln I(x) \rangle}$ decays faster than the average intensity $\langle I(x) \rangle$.

For inhomogeneous random media, the separation between neighboring scatterers follows a L\'evy $\alpha$-stable distribution with support on positive real numbers. 
Because the disorder is inhomogeneous, the method of building a disordered system is important. We build these disordered structures by sequentially adding scatterers starting from the left side. 
As a result, the density of scatterers is higher on the left of the lattice and gradually decreases toward the right end. This process introduces a significant probability of large separations between scatterers, as dictated by a L\'evy stable distribution. For inhomogeneous disorder modeled by a L\'evy distribution with $\alpha < 1$, we obtain a power-law relationship for the logarithmic intensity, on both sides of the position $x_0$:
\begin{equation}
\label{lnI_Levy}
\langle \ln I(x) \rangle = \left\{
\begin{array}{ll}
\langle \ln T \rangle_{0,x_0} \left[1- \left(\frac{x}{x_0}  \right)^\alpha \right]  & \textrm{, for $ x < x_0$}  \\ \\
 \langle \ln T \rangle_{x_0,L} \left(\frac{ x-x_0}{L-x_0}  \right)^\alpha & 
\textrm{, for $ x > x_0$}
\end{array} \right.
\end{equation}
where $\langle \ln T \rangle_{0, x_0}$ and $\langle \ln T \rangle_{x_0, L}$ are the averages of the logarithmic transmission of the segments of length $x_0$ and $L-x_0$, at the left and the right of $x_0$, respectively (see more details in~\cite{SM}).  

As we remarked previously, for homogeneous media, the typical intensity behaves as the average intensity, but it exhibits a different decay rate. 
We extend this result to inhomogeneous disorder by conjecturing 
that the average intensity decays according to a stretched exponential function for large distances $|x-x_0|$: 
\begin{equation}
\label{I_Levy}
\langle  I(x) \rangle \sim\left\{
\begin{array}{ll}
 e^{-\lambda_1 \left[1- \left(\frac{x}{x_0}  \right)^\alpha \right]}  & \textrm{, for $ x < x_0$}  \\ \\
 e^{-\lambda_2 \left(\frac{ x-x_0}{L-x_0}  \right)^\alpha } & 
\textrm{, for $ x > x_0$}
\end{array} \right.
\end{equation}
where $\lambda_1$ and $\lambda_2$ are constants.  We note that the average intensity decreases more slowly with distance compared to the standard Anderson localization described by Eq.~(\ref{I_Anderson}). Therefore, the previous results, Eqs. (\ref{lnI_Levy}) and (\ref{I_Levy}) show the significant effects of the inhomogeneous disorder on wave diffusion.

We conducted numerical simulations to support our theoretical predictions and experimental results, discussed below. Since photonic lattices are composed of well-confined waveguides with a single bound state, the propagation of electromagnetic (EM) waves can be modeled by coupled mode theory, which perfectly agrees with the tight-binding approximation from solid-state physics~\cite{D_yerville_2002,Lederer_2008,Flach_2008,Vicencio_2021}. 
\begin{figure}
\centering
\includegraphics[width=1.\columnwidth]{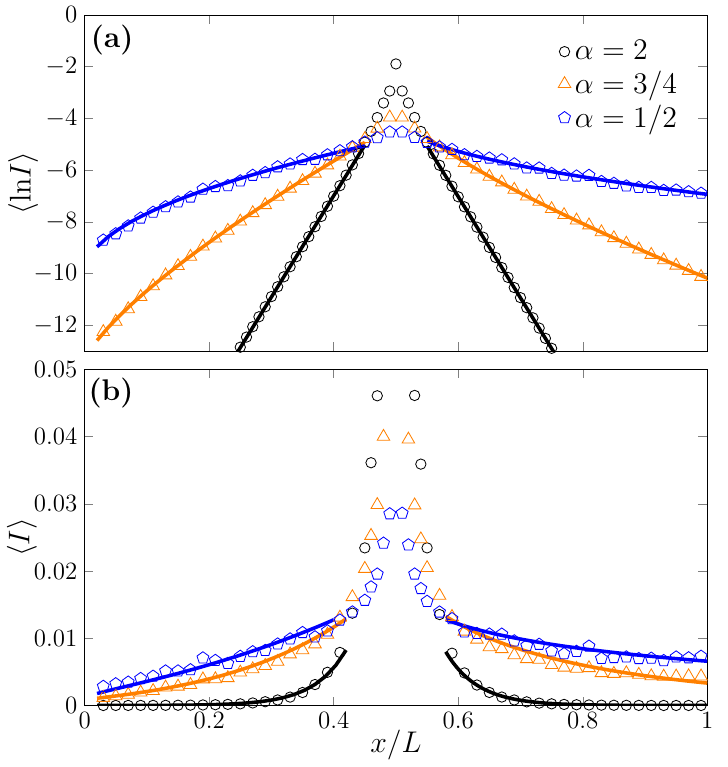}
\caption{
    (a) Theoretical and numerical results for the averaged logarithmic intensity and (b) the averaged intensity, for different values of $\alpha$, as indicated in (a). Colored symbols (pentagons, triangles and circles) show the numerical results, while solid lines of the same color show the corresponding theoretical predictions (\ref{lnI_Anderson}) and (\ref{lnI_Levy}) in panel (a) and (\ref{I_Anderson}) and (\ref{I_Levy}) in panel (b).}
    \label{fig2}
\end{figure} 
This leads us to the Hamiltonian
\begin{equation}
H = \sum_{j=1}^L \varepsilon_j c^\dagger_j c_j- \tau\left( c^\dagger_j c_{j+1}  + c^\dagger_{j+1} c_j \right)\ ,\label{model}
\end{equation}
where $c_j^\dagger$ and $c_j$ are the respective creation and annihilation operators acting on site $j$. $\tau$ is the nearest-neighbor hopping amplitude, which we fix to $\tau=1$, and $L$ is the number of lattice sites. In our problem, the on-site energy $\varepsilon_j$ can take two values: $0$ or $W$. The sites with energy $W$ are randomly separated according to a L\'evy $\alpha$-stable distribution, which is numerically generated as explained in~\cite{SM}. 
The assignment of sites with on-site energy $W$ starts on the left side of the chain, leading to a higher density of sites with energy $W$ on the left side of the lattice.

We consider 1D chains with $L=100$ sites and disorder strength $W=3$ in our numerical simulations. We average over $110000$ independent realizations, where for each configuration, a delta wavepacket is released at site $x_0$. The wavepacket evolves over a dimensionless time ($\hbar = 1$) up to $t=500$. As we have numerically verified, this evolution is sufficient for the diffusion process to cease and for the average intensity to reach a stationary profile. As a result, we obtain the intensity statistics as a function of the position $x$. The results of the numerical simulations and their comparison with the theoretical predictions summarized by Eqs.~(\ref{lnI_Anderson})-(\ref{I_Levy}), for homogeneous and inhomogeneous disorder, are shown in Fig.~\ref{fig2}. In all the considered cases, the initial wave packet was released in the center of the samples at $x_0/L = 1/2$ (we provide results for different release positions in the Supplemental Material~\cite{SM}). For homogeneous disorder, the numerical ensemble average $\langle \ln I(x) \rangle$ is presented in Fig.~\ref{fig2}(a) using black symbols. The numerical simulations show a linear behavior of $\langle \ln I(x) \rangle$, as described by  Eq.~(\ref{lnI_Anderson}). We also notice that $\langle \ln I(x) \rangle$ exhibits a symmetric decay from the position where the wavepacket was injected.

For inhomogeneous disorder, characterized by $\alpha=1/2$ and $3/4$, the numerical results for $\langle \ln I(x) \rangle$ are shown in Fig.~\ref{fig2}(a) with blue and orange symbols, respectively. A power-law decay is found. The solid lines in Fig.~\ref{fig2}(a) are fits of Eq.~(\ref{lnI_Levy}) using a single additive constant as a fitting parameter for both the left and right sides of the averaged profile, relative to the excitation position. In contrast to homogeneous disorder, we clearly notice that for an inhomogeneous case $\langle  \ln I(x) \rangle$ exhibits an asymmetric decay from the wavepacket's release position $x_0$.

The numerical results for the average intensity $\langle I(x) \rangle$, for homogeneous and inhomogeneous disorder, are shown in Fig.~\ref{fig2}(b). We observe exponential wave localization for homogeneous disordered structures (see black symbols), as expected from the prediction~(\ref{I_Anderson}) shown by black solid lines. On the other hand, for inhomogeneous disordered systems, characterized by $\alpha=1/2$ and $3/4$, the wave localization is described by stretched exponential functions, as given by  Eqs.~(\ref{I_Levy}), which is in contrast to the widely known exponential Anderson localization. The blue and orange solid lines in Fig.~\ref{fig2}(b) correspond to fits of the tails distributions based on Eq.~(\ref{I_Levy}), with $\lambda_1$ and $\lambda_2$ as fitting parameters. Again, the simulations and the theoretical results reveal an asymmetric decay from the input position, which stands in contrast to the symmetric decay found for homogeneous disordered systems.
\begin{figure}
\centering
\includegraphics[width=1.\columnwidth]{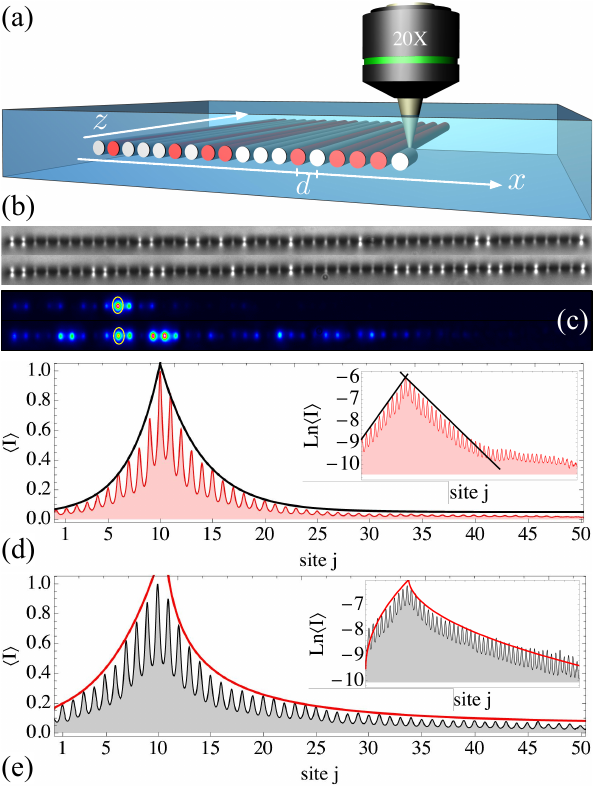}
   \caption{(a) Sketch of the femtosecond laser writing technique, where light-gray and red disks represent waveguides with $\varepsilon_j=0$ and $W$, respectively. (b) White light microscope images for two fabricated lattices with $L=50$ sites each. (c) Intensity output images for the excitation of site $j=10$ (yellow circles) for $\alpha=2$ (top) and $\alpha=1/2$ (bottom). (d) and (e) Averaged intensity for $\alpha=2$ and $\alpha=1/2$ distributions. The envelopes represented by solid lines are fits based on Eqs.~(\ref{I_Anderson}) and (\ref{I_Levy}). The insets in these panels show the logarithmic plots with the linear and power-law decays, consistent with the predictions from Eqs.~(\ref{lnI_Anderson}) and (\ref{lnI_Levy}), represented by solid lines.}  
\label{fig3}
\end{figure}

We experimentally implement the tight-binding model represented by Eq.~(\ref{model}) by fabricating several photonic lattices using the femtosecond laser-written technique~\cite{Szameit_2005}, as it is sketched in Fig.~\ref{fig3}(a). This technique allows the fabrication of a photonic lattice formed by a set of single-mode waveguides on a defined geometry. The waveguides length defines the propagation distance along $z$ [see Fig.~\ref{fig3}(a)], which acts as a dynamical analog of time in tight-binding-like systems (see Ref.~\cite{SM} for more details). The coupling constant (hopping) $\tau$ decays exponentially with the inter-site distance $d$~\cite{Guzm_n_Silva_2021}, which also defines the dynamical range for the system. For this specific work, we are required to have the largest effective propagation distance but also strong on-site energies $W$. When normalizing the dynamical equations~\cite{SM}, we realize that the effective propagation distance is redefined as $z_e\equiv \tau z$, while the on-site energies are given as $\varepsilon_j/\tau$. Therefore, once we experimentally increase $\tau$, by, for example, decreasing the waveguide separation $d$, we increase $z_e$ and simultaneously decrease the effective on-site energies.

We fabricate lattices with waveguides, setting the on-site energies to $0$ and $W$. Inhomogeneous disorder is introduced by randomly inserting waveguides $W$ according to an $\alpha$-stable distribution with $\alpha=1/2$.  For the case of homogeneous disorder,  we use a Gaussian distribution, $\alpha=2$. All the samples were fabricated on a $10$ cm long borosilicate glass wafer, as shown in the examples of Fig.~\ref{fig3}(b), where the brighter sites correspond to on-site energies $W$. Each lattice has an inter-site distance $d=16\ \mu$m and a total number of $L=50$ sites~\cite{SM}. 
As a fabrication routine, we fix the fabrication power to $30$ mW and vary the writing velocity $v$ to fabricate different waveguides (the slower the velocity $v$, the stronger the waveguide). Therefore, nominal waveguides ($\varepsilon_j=0$) are fabricated at $v=20$ mm/s, while the stronger waveguides ($\varepsilon_j=W$) at $v=3$ mm/s. We optimized the excitation wavelength to $730$ nm to achieve a good balance between transport and localization for various disorder realizations (a larger wavelength produces wider waveguide modes that couple more strongly with their nearest-neighbors~\cite{C_ceres_Aravena_2023}). The excitation site $x_0$ was set at $j_0=10$, and we also excited $10$ additional sites around $j_0$ to increase the experimental statistics. Thus, the experimental ensemble was composed of 418 images for each disorder case. We estimate the experimental parameters as $\tau\approx 1$ cm$^{-1}$, $W\approx 3$ cm$^{-1}$, and $z_{e}\approx 25$ cm~\cite{SM}. Fig.~\ref{fig3}(c) shows two examples of single measurements after exciting the site $j=10$ for $\alpha=1/2$ (top) and $\alpha=2$ (bottom). Additionally, the case for $\alpha=3/4$ is shown in~\cite{SM}.

After measuring all the fabricated lattices, we analyzed the data by projecting the experimental profiles on the horizontal axis. This allowed us to obtain the site's intensities for all the images, and we shifted the positions so that all of them were centered at the site $j=10$, as the excitation site. Following this adjustment, we obtained the averaged intensity profiles shown in Figs.~\ref{fig3}(d) and (e) for Anderson and anomalous localization, respectively. Additionally, we include the logarithm of the average intensity as insets. We observe a significant contrast between the two cases. For homogeneous disorder, there is a very well-defined exponential-averaged profile that decays very fast into the lattice, as shown by the black curves fitting the average envelope. This aligns with the typical scenario of Anderson's localization and, accordingly, the inset in Fig.~\ref{fig3}(d) demonstrates a linear dependence away from the excitation site, as it is well-fitted by the straight lines. However, the case of inhomogeneous disorder $\alpha=1/2$ produces a noticeably different result, which is characterized by a slower decay into the lattice, as shown in Fig.~\ref{fig3}(e). The profiles in Fig.~\ref{fig3}(e) are in excellent agreement with the prediction from Eq.~(\ref{I_Levy}), shown by red curves, and the numerical results described in Fig.~\ref{fig2}. The logarithm of the average intensity shows a very contrasting profile compared to the Anderson case, with a very well-fitted profile according to Eq.~(\ref{lnI_Levy}), and with a characteristic asymmetric decaying tail.

In conclusion, we investigated wave localization phenomena in inhomogeneously disordered one-dimensional systems by measuring the output light intensity of disordered photonic lattices. We introduce disorder into the lattices by incorporating waveguides with onsite energies $W$ randomly separated according to a heavy-tailed distribution known as the $\alpha$-stable or L\'evy stable distributions. 
Our findings demonstrate that the power-law tails of the Levy stable distributions play a fundamental role in the wave statistics within the photonic lattices. We discovered that light exhibits an anomalous localization in relation to the standard exponential decay in Anderson's phenomenon. Instead of this standard decay,  the weaker localization we observed is characterized by a stretched exponential decay with a stretching parameter $\alpha$.

Our theoretical model indicates that both the average logarithmic transmission of the left and right segments with respect to $x_0$ and the stretching parameter $\alpha$ completely describe the wave statistics. Additionally, we observe that light localization is not symmetric with respect to the input position. From a practical point of view, the phenomenon of localization has been exploited to enhance the transport of images in optical fibers and the transport of photons in photonic crystals~\cite{Mosk2012,Hsieh2015}. Therefore, our experimental results and theoretical framework could pave the way for further engineering and control of wave diffusion in random media. 

Slowly decaying intensity profiles, such as those described by a stretched exponential function, could provide an interesting approach for studying long-range effects in physics, with enhanced interactions among distant lattice regions or even different systems. For example, standard Kerr-like nonlinear interactions are very local and have no effect away from the respective site, making the dynamics relatively simple 
~\cite{Flach_2008}. However, incorporating long-range and saturation effects~\cite{Lederer_2008} could promote dynamical transitions and more complexity, where slow decaying intensities could be a good starting point for exploration. Additionally, studying disorder in finite lattices is an important issue of practical interest and an avenue for future research. Most theoretical studies assume infinite lattices and nearest-neighbor interactions, which may not be valid for large systems, where weak coupling coefficients coming from next-nearest-neighbours can play a significant role. Consequently, exploring real, finite disordered systems with shorter dynamical ranges is an interesting path for future experimental investigations.

A.R.Y. and T.G. acknowledge financial support from CONAHCYT (México) under project number CF-2019 10872. R.A.V acknowledges financial support from Millennium Science Initiative Program ICN17$\_$012 and FONDECYT Grant No. 1231313. V.A.G. recognizes financial support by MCIU (Spain) under the Project number PGC2018-094684-B-C22. 

\bibliography{Levy_Photonic_Bib}
\end{document}


\title{Supplemental Material for ``Anomalous localization of light in one-dimensional L\'evy photonic lattices''}

\author{Alejandro Ram\'irez-Yañez}

\affiliation{Departamento de F\'isica, CUCEI, Universidad de Guadalajara, Guadalajara, Jalisco, C.P. 44430, México}

\author{Thomas Gorin}

\affiliation{Departamento de F\'isica, CUCEI, Universidad de Guadalajara, Guadalajara, Jalisco, C.P. 44430, México}

\affiliation{Max-Planck-Institut für Physik komplexer Systeme, Nöthnitzer Str. 38, D-01187 Dresden, Germany\looseness=-1}

\author{Rodrigo A. Vicencio}

\affiliation{Departamento de F\'isica, Facultad de Ciencias F\'isicas y Matem\'aticas, Universidad de Chile, Santiago 8370448, Chile}

\affiliation{Millenium Institute for Research in Optics–MIRO, Universidad de Chile, Chile}

\author{V\'ictor A. Gopar}

\affiliation{Departamento de F\'isica Te\'orica and BIFI, Universidad de Zaragoza, Pedro Cerbuna 12, E-50009, Zaragoza}

\maketitle
\makeatletter
\renewcommand{\theequation}{S\arabic{equation}}
\renewcommand{\thefigure}{S\arabic{figure}}
\renewcommand{\bibnumfmt}[1]{[S#1]}
\renewcommand{\citenumfont}[1]{S#1}
\vspace{-0.5 cm}


\section{Ensemble average of the logarithmic intensity in 1D  random media} 

\subsection{Inhomogeneous disorder}

In this section, we derive the average of the logarithmic intensity 
$\langle \ln I(x) \rangle$ as given by Eq.(4) in the main text. The derivation follows the results in Refs.~\cite{Razo_Lopez_2021, Falceto_2010}.

In Ref.~\cite{Falceto_2010}, it was found that for structures of length $L$ with disorder modeled by L\'evy stable distributions, the probability density of the number of scatterers $\Pi_{L}(n;\alpha)$ in a system of length $L$ is given by
\begin{equation}
\label{pi_alpha}
 \Pi_{L}(n;\alpha)=\frac2\alpha\frac{L}{(2n)^{\frac{1+\alpha}{\alpha}}}
q_{\alpha,c}\left({L}/{(2n)^{1/\alpha}}\right) ,
\end{equation}
for $0 < \alpha <1 $, with $c$ a scaling parameter, in the limit $L\gg c^{1/\alpha}$. The probability density $q_{\alpha,c}(x)$ in Eq.~(\ref{pi_alpha}) is a L\'evy stable distribution with parameter $\alpha$. Thus, for $x \gg 1$, $q_{\alpha,c}(x)$ has a power-law tail: $q_{\alpha,c}(x) \sim c/x^{1+\alpha}$. It was assumed in Ref.~\cite{Razo_Lopez_2021} that the 
the disordered structures were constructed by adding scatterers starting from the left end at $x=0$.

\begin{figure}[H]
  \center{\includegraphics[width=0.6\columnwidth]{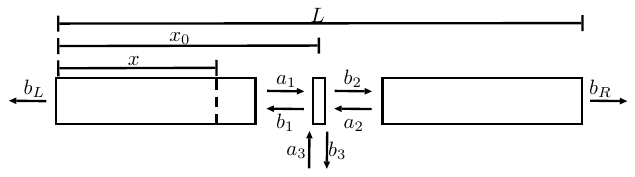}}
\caption{Model of 1D chain with transmitted and reflected amplitude coefficients $a_i$,$b_i$ in each scatterer. $x_0$ is the releasing point where waves start to spread out (the injection point), $x$ is the measuring point, $L$ is the system length.} 
\label{logI_ScatteringProblem}
\end{figure}

Within a random matrix theory framework to scattering, it was shown in Ref.~\cite{Razo_Lopez_2021} that for incident waves arriving from the left end of structures, the average of the logarithmic intensity at a point $x$ can be given in terms of the average of the logarithmic transmissions as $\langle \ln I(x)\rangle= \langle \ln T \rangle_{0,L} -\langle \ln T \rangle_{x,L}$.  That is, $\langle \ln I(x)\rangle$ is given by the difference of the average of the logarithmic transmissions of the total sample of length $L$, $\langle \ln T \rangle_{0,L}$, and the average of the logarithmic transmission of the right segment to the observation point $x$, $\langle \ln T \rangle_{x,L}$. Thus, using that the logarithmic transmission is an additive quantity and  performing the ensemble averages using Eq.~(\ref{pi_alpha}), it was found that $\langle \ln I(x)\rangle$ can be expressed as
\begin{equation}
   \langle \ln I(x) \rangle = \langle{\rm{ln}} T \rangle_{0,L}\left(\frac{x}{L}\right)^\alpha ,
\end{equation}
where $\langle{\rm{ln}}(T)\rangle_{0,L}=-\frac{a}{c}\mathcal{I_\alpha}L^\alpha$ with $a$ a constant, and $\mathcal{I_\alpha}=\frac{1}{2}\int {\rm{d}}z~z^{-\alpha}q_{\alpha,1}$.
We thus apply the above analysis to the system depicted in Fig.~\ref{logI_ScatteringProblem}.

(a) First, let us consider the left segment to the node at $x_0$ of the 1D structure. In this case, the logarithmic intensity is given by the difference
\begin{equation}
   \langle \ln I(x) \rangle =\langle{\ln}T\rangle_{0,x_0}-\langle{\rm{ln}} T \rangle_{0,x},  \\
   \end{equation}
which can be written as
   \begin{equation}
     \langle \ln I(x) \rangle = \langle{\ln}T\rangle_{0,x_0}\left[ 1 - \left( \frac{x}{x_0}\right)^\alpha  \right]
     \nonumber \label{Intensity_left}, 
\end{equation}
namely, Eq.(4) for $x < x_0$.

(b) For the right segment to the node at $x_0$, we make $x \to x-x_0$, and the logarithmic intensity is thus given by the difference
\begin{align}
   \langle \ln I(x) \rangle &=\langle{\ln} T\rangle_{L,x_0}-\langle{\rm{ln}} T\rangle_{L,x} \nonumber \\
      \label{Intensity_right}
\end{align}
If we write the segment of length $L-x$ as $L-x= (L-x_0)- (x-x_0)$, then, from Eq.~(\ref{Intensity_right}), we write $ \langle \ln I(x) \rangle =\langle{\ln}T\rangle_{x,x_0}$, which can be expressed as 
\begin{align}
   \langle \ln I(x) \rangle &=\langle{\ln}T\rangle_{x_0,L} \left( \frac{x-x_0}{L-x_0} \right)^\alpha  \nonumber
\end{align}
namely, Eq.(4) for $x>x_0$.

\subsection{ Homogeneous disorder}

The average logarithmic intensity for the case of homogeneous disorder is obtained following the same steps as for the previous case of inhomogeneous disorder, with the difference that $\langle \ln I(x) \rangle$ is a linear function of the position $x$. For incident waves from the left of the samples, it was found that~\cite{Cheng_2017}:
\begin{equation}
   \langle \ln I(x) \rangle = \langle{\rm{ln}} T\rangle_{0,L}\left(\frac{x}{L}\right) .
\end{equation}
with $\langle \ln T \rangle_{0,L} = - L/\ell$ where $\ell$ is the mean free path.  Thus, from Eq. (S3), for the left segment to the node at $x_0$: $\langle \ln I(x) \rangle = -( x_0 -  x)/\ell $, while for the right segment to the node, from Eq.(S4), $\langle \ln I(x) \rangle = -( x -  x_0)/\ell $.  These two results lead to Eq. (2) in the main text:
\begin{equation}
   \langle \ln I(x) \rangle = -\left| x-x_0 \right|/\ell . \nonumber
\end{equation}

\section{Average logarithmic intensity for various positions of the excitation point and different initial states }

\begin{figure}[H]
  \center{\includegraphics[width=0.6\columnwidth]{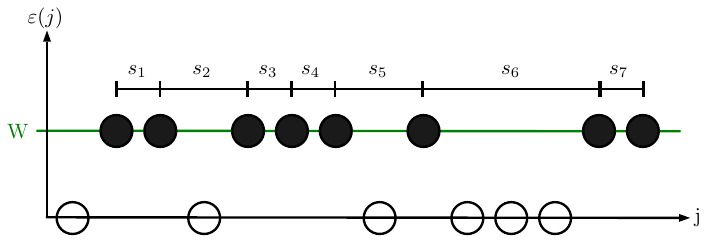}}
\caption{Schematic of a disordered 1D chain with randomly separated sites. The random distances $s_j$ follow a L\'evy stable distribution. Black dots represent sites with energy $\varepsilon=$ W, while empty dots stand for sites with $\varepsilon=0$.} 
\label{DistancesConstruct}
\end{figure}

The numerical simulations for ensemble averages of the intensity $I(x) = |\Psi(x)|^2$ shown in Fig. 2 in the main text, were conducted by allowing an initial wave packet $\Psi_0$ to evolve according to the tight-binding Hamiltonian that describes the dynamics of 1D disordered chain $H$, Eq. (6). A schematic of the 1D chain is shown in Fig~\ref{DistancesConstruct}.

We note that the Hamiltonian, Eq. (6), models single-particle systems where the number of particles remains constant. The evolution of the wave packet is thus obtained by propagating $\Psi_0$: $\Psi(t)=e^{itH/\hbar}\Psi_0$. For each disorder configuration, the initial wave packet is released at position $x_0$ and evolves over time until the ensemble average intensity reaches a stationary profile, i.e., until wave diffusion ceases. 

In the main text, we have presented the results of the ensemble average $\langle \ln I(x) \rangle$  for homogeneous and inhomogeneous 1D random media for the particular case of the excitation point at $x_0/L =1/2$ and a Dirac delta wave packet as the initial state.

In this section, we show the results for different positions of the excitation point $x_0$ for inhomogeneous disorder modeled by $\alpha$-stable distributions with $\alpha=1/2$ and 3/4 and homogeneous disorder modeled by a Gaussian distribution ($\alpha=2)$. Additionally, we present the results of 
the average $\langle \ln I(x) \rangle$ when the initial state is Gaussian instead of a Dirac delta.

\subsection{Dirac delta initial state released at different  positions}

\begin{figure}[H]
\center{\includegraphics[width=0.9\columnwidth]{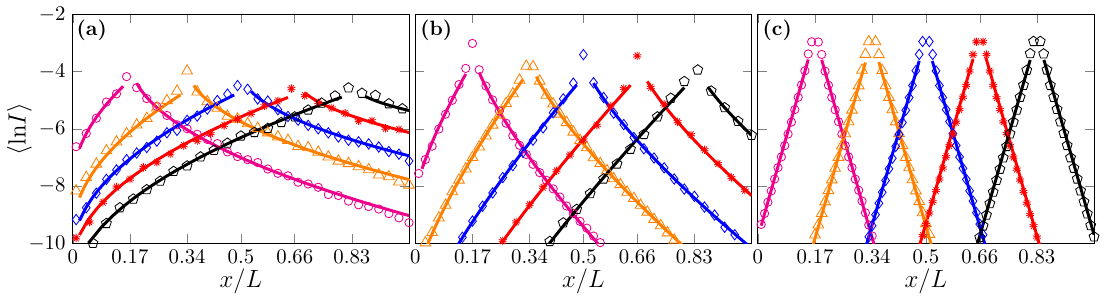}}
\caption{The average of the logarithmic intensity for different positions: $x_0/L=1/6,1/3, 1/2,4/5$, and 5/6 of the initial Dirac delta wave packet. 
Symbols represent the numerical results, while the theoretical predictions, Eq. (4), are shown as solid lines. Panels (a), (b), and (c) stand for $\alpha=1/2$, 3/4, and 2, respectively. 
Localization is observed for all different initial positions of a delta wave packet propagated on time.}
\label{DifferentPositions}
\end{figure}

The results for five different initial positions $x_0/L$ using Dirac delta wave packets as initial state are shown in Fig.~\ref{DifferentPositions}. In all the cases presented in Fig.~\ref{DifferentPositions} (a) and \ref{DifferentPositions} (b), a power-law decay of $\langle \ln I(x) \rangle$ is obtained and is well described by the theoretical results, Eq. (4).

\subsection{Gaussian initial state}

A natural question is to ask how the intensity statistics are affected by the landscape of the initial wave packet. Thus, instead of a Dirac delta wavefunction, we use a more realistic Gaussian wave packet $\Psi_0=A{\exp}\left[-{(x-x_0)^2}/{2\sigma_0^2}\right]$, where $A$ is a normalization coefficient, as the initial state.

\begin{figure}[H]
  \center{\includegraphics[width=0.9\columnwidth]{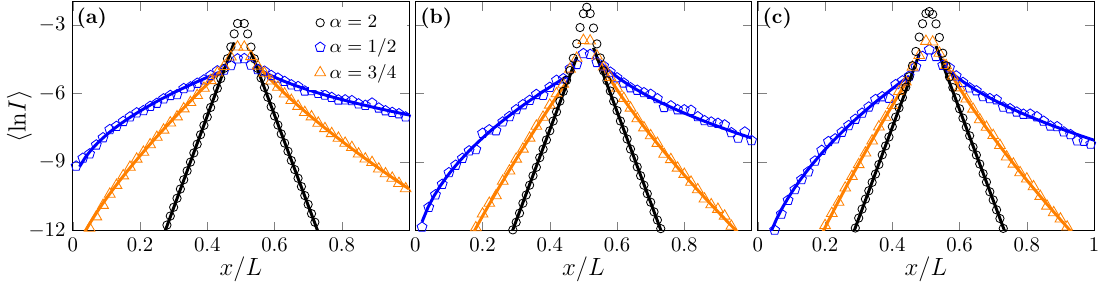}}
\caption{The average of the logarithmic intensity for  
different widths, $\sigma_0$, of Gaussian initial packets: $\sigma_0=2$ in panel (a) and 
$\sigma_0=\sqrt{2}$ in panel (b). Additionally, we show in panel (c) the case for delta initial states $\Psi_0=\delta(x-x_0)$. Symbols and solid lines represent the numerical and theoretical results, respectively. Anomalous localization is observed for all different initial states propagated on time for $\alpha= 1/2$ and $3/4$. Anderson localization is observed as well for all initial states for $\alpha=2$}
\label{GaussianInitStates}
\end{figure}

In Figs.~\ref{GaussianInitStates} (a) and \ref{GaussianInitStates} (b), we show the results for two different widths $\sigma_0$ of the Gaussian wave packets for inhomogeneous ($\alpha=1/2, 3/4$) and homogeneous ($\alpha=2$) disorders. Additionally, Fig.~\ref{GaussianInitStates}(c) displays the results for Dirac delta wave packets as an initial state for comparison with the cases in panels \ref{GaussianInitStates} (a) and \ref{GaussianInitStates} (b). We can observe a power-law decay of $\langle \ln I(x) \rangle$ for L\'evy disordered structures, whereas the decay is linear for the homogeneous disorder. It can be seen from Fig.~\ref{GaussianInitStates} that the theoretical results in Eq. (4) (solid lines) describe the numerical results.

\section{Anomalous localization in inhomogeneous media with different profiles of the density of scatterers}

In the main text, we focus on the case where the construction of the 1D disordered chains begins from the left end of the structure. This way of constructing the 1D chain results in a higher density of sites with energy $W$ in the left part of the structure.

In this section, we examine the intensity statistics of two different scenarios: 
(a) when the construction begins from the right end, and (b) when it starts from the center of the chain extending towards both the left and right sides. 

\begin{figure}[H]
  \center{\includegraphics[width=0.7\columnwidth]{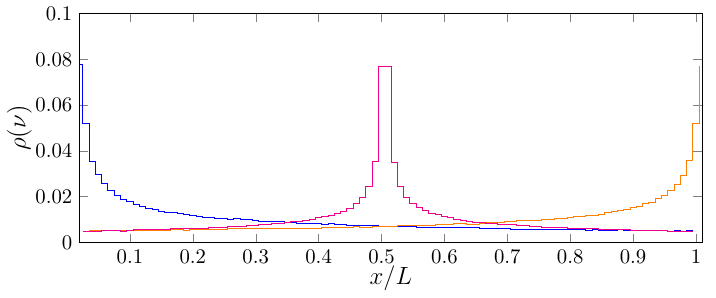}}
\caption{Scatterer density $\rho(\nu)$ resulting from the different chain construction schemes.  The histogram in blue (orange) represents the density of chains constructed beginning from the left (right) end, while the histogram in magenta   
corresponds to the density of chain construction beginning from the center of the samples.} 
\label{ScattererDensity}
\end{figure}

In Fig.~\ref{ScattererDensity}, we illustrate the landscape of the density of sites $\rho(\nu)$ with energy $W$ for the cases (a) (orange histogram)  and (b) (magenta histogram), as indicated above. The density $\rho(\nu)$ for the case considered in the main text is also shown (blue histogram).  

\begin{figure}[H]
  \center{\includegraphics[width=0.9\columnwidth]{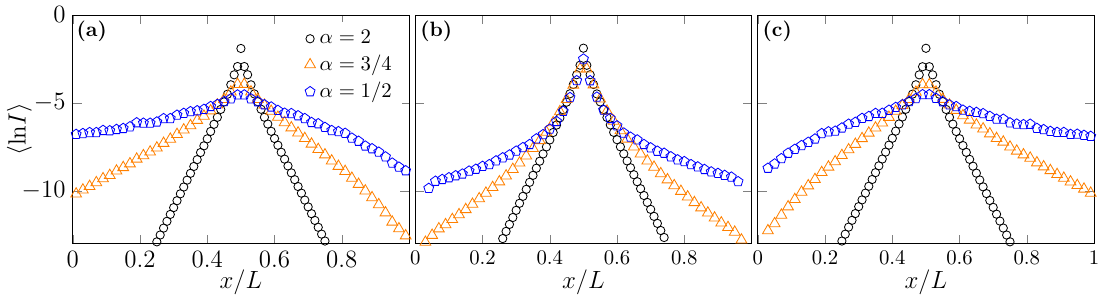}}
\caption{Average of the logarithmic intensity for the schemes of chain construction. (a) construction begins from the right end, (b) from the center, and (c) from the left end; this case corresponds to the one used in the main text.} 
\label{lnI_DiffSchemes}
\end{figure}

We thus obtain the ensemble average of the logarithm intensity for the construction cases (a) and (b). The results are presented in Fig~\ref{lnI_DiffSchemes}.   
For comparison, in Fig.~\ref{lnI_DiffSchemes}(c), we also include the case discussed in the main text, where the construction starts from the left.

In Fig.~\ref{lnI_DiffSchemes} (b), we observe a symmetric decay of $\langle \ln I(x) \rangle$ respect to the excitation point $x_0/L =1/2$. In contrast, $\langle \ln I(x) \rangle$ in panel (a) exhibits a nonsymmetric decay, and as expected, it corresponds to the vertical mirror image of the case presented in panel (c).

\section{Simulation and implementation of random numbers from a $\alpha$-stable distribution}

\textit{Simulation.} Simulating a set of random numbers from a stable distribution is not straightforward due to the non-existence (except for some special cases) of the inverse $\mathbf{F}^{-1}(x)$ nor the CDF $\mathbf{F}(x)$, such as the usual approaches as the inversion method are not a viable solution; other algorithms as the rejection method would require tedious computations. Instead, Chambers, Mallows, and Stuck proposed an efficient solution in the so-called CMS algorithm Ref.~\cite{Chambers_1976}.\\

There are four parameters describing a stable distribution: the stability parameter $0 < \alpha \leq 2$, the asymmetry parameter $-1\leq \beta \leq 1$, the location parameter $\mu \in \mathbb{R}$, and the scale parameter $c \geq 0$, plus a hyper-parameter describing its parametrization; different parametrizations are useful for different goals. Nolan \cite{Nolan_2020} describes more than ten different parameterizations of $\alpha$-stable distributions, but two are the most commonly used, labeled $k=0$ and $k=1$.\\

We used the definitions given in Ref.~\cite{Nolan_2020}: A random variable $X$ is  of the class $\mathbf{S}(\alpha,\beta,\mu,c;k=0)$ if
\begin{align}
    X\overset{\mathrm{d}}{=}
    \begin{cases}
      c(Z-\beta{\rm{tan}}(\frac{\pi \alpha}{2}))+\mu \qquad &\alpha \neq 1,\\
      cZ+\mu \qquad &\alpha=1.
    \end{cases}
\end{align}
where $\overset{\mathrm{d}}{=}$ means equality in distribution, i.e. both expressions have the same probability law. The random variable $Z$ is defined below, following the CMS algorithm. Let $\Theta$ and $W$ be independent with $\Theta$ uniformly distributed on ($-\frac{\pi}{2},\frac{\pi}{2}$), $W$ exponentially distributed with mean $1$. Thus,  the symmetric random variable
\begin{align}
    Z=
    \begin{cases}
        \frac{{\rm{sin}}(\alpha\Theta)}{{\rm{cos}}(\Theta)^{1/\alpha}}\left[ \frac{{\rm{cos}}((\alpha-1)\Theta)}{W}\right]^{(1-\alpha)/\alpha} \qquad &\alpha \neq 1,\\
        {\rm{tan}}(\Theta) \qquad &\alpha=1,
    \end{cases}
\end{align}
has a $\mathbf{S}(\alpha,\beta=0,\mu,c;k=0)$ $=$ $\mathbf{S}(\alpha,\beta=0,\mu,c;k=1)$ distribution.\\

\begin{figure}[H]
\center{\includegraphics[width=0.7\columnwidth]{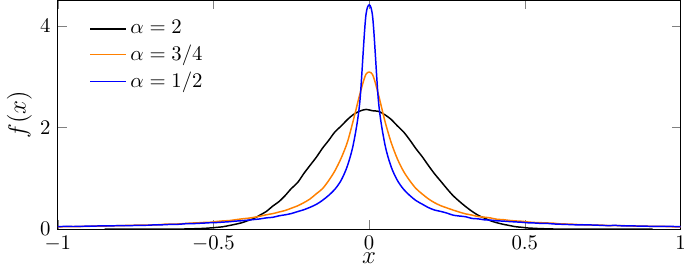}}
\caption{Probability density functions for different values of $\alpha$ using the parametrization $\mathbf{S}(\alpha,\beta=0,\mu,c;k=0)$ with $\mu=0$. The distributions are symmetric around $\mu=0$.} 
\label{PDFs_Symmetric}
\end{figure}

\textit{Implementation.} We want to generate ensembles of discrete 1D chains consisting of $N$ sites with energies $\varepsilon_i=0$ and $W$. The sites with energies $W$ are separated by distances taken as random numbers from a L\'evy stable distribution. In the simplest case, the distribution is symmetric about $\mu$. The stable distribution corresponding to $\alpha=2$ is a Gaussian distribution in any class representation Ref.~\cite{Nolan_1997}, see figure(\ref{PDFs_Symmetric})). For convenience, we use $\mu=0$. This is the case used to observe Anderson's localization. We choose the class of distributions $\mathbf{S}(\alpha,\beta=0,\mu=0,c;k=0)$ with support in $\mathbb{R}$, symmetric and centered with $\mu=0$ to generate our distances for the 1D discrete chains and count only positive members of the set as valid distances.

In the context of the wave scattering problem, the tails of the distributions are the most important feature and are determined only by the stability parameter $\alpha$; for instance,  see Ref.~\cite{Falceto_2010} for more details. For all results shown in the main text, the distribution classes $\mathbf{S}(\alpha,\beta=0,\mu,c;k=0)$ were used to construct the chains.

\section{Dynamical normalization}

A generic Hamiltonian for a one-dimensional (1D) lattice can be written as in Eq. (6):
\begin{equation}
H = \sum_{j=1}^N \left[\varepsilon_j c^\dagger_j c_j+ \tau\left( c^\dagger_j c_{j+1}  + c^\dagger_{j+1} c_j \right)\right],\nonumber
\label{model}
\end{equation}
In a classical representation, this Hamiltonian, obtained from coupled mode theory~\cite{CMTOkamoto}, can be directly written as
\begin{equation}
H = \sum_{j=1}^N \left[\varepsilon_j |c_j|^2+ \tau \left(c_{j+1} c^*_j + c^*_{j+1} c_j \right)\right].\label{model}
\end{equation}
In the photonics context $\varepsilon_j$ describes the propagation constant of the $j$-th waveguide, with the respective propagation coordinate $z$ as the dynamical variable, and $\tau$ corresponds to the coupling constant between neighboring waveguides. The dynamical equations are obtained by differentiating $H$ with respect to $c_j^*$, obtaining  
\begin{equation}
    -i\frac{\partial c_j}{\partial z}=\frac{\partial H}{\partial c_j^*}= \varepsilon_j c_j+\tau(c_{j+1}+c_{j-1})\ .
\end{equation}
This is the well-known Discrete Linear Schr\"odinger (DLS) equation~\cite{Lederer_2008}, which describes the evolution of an electromagnetic wavefunction $c_j$ on a tight-binding-like 1D photonic lattice. We can always normalize this equation, and eliminate their respective units, by simply dividing by ``$\tau$'', obtaining
\begin{equation}
    -i\frac{\partial c_j}{\partial \xi}= \bar{\varepsilon}_j c_j+(c_{j+1}+c_{j-1})\ , \label{DLS}
\end{equation}
where $\bar{\varepsilon}_j\equiv \varepsilon_j/\tau$ and $\xi\equiv \tau z$. We immediately notice that the effective dynamical variable is indeed $\xi$ and, experimentally, this can be modified by taking a larger glass such that we can have a larger propagation distance, or by simply increasing the coupling constant $\tau$. The coupling constant depends essentially on two parameters: the waveguide distance (the farther the waveguides, the smaller the coupling) and the mode wavefunctions (the wider the modes, the larger the coupling)~\cite{Szameit_2005}. Therefore, a way of directly increasing the hopping is by modifying the guided mode. This can be done by adjusting the fabrication parameters to have stronger or weaker guiding structures. But, once the waveguide is created, a way of changing the mode profile is by adjusting the polarization~\cite{Rojas_Rojas_2014} or the excitation wavelength Ref.~\cite{C_ceres_Aravena_2023}. The first method has only two independent possibilities, while the last method is much more efficient and can be easily implemented by means of a supercontinuum laser source: the larger the wavelength, the larger the mode width. Therefore, a sweep in the excitation wavelength can directly be associated with ramping in the coupling constant and, therefore, with an increment of the effective dynamical variable $\xi$.

\section{Femtosecond laser writing technique and initial characterization}

We fabricate different photonic structures by using a femtosecond (fs) laser writing technique~\cite{Szameit_2005}, as it is sketched in Fig.2(a) of the main text. Ultrashort pulses from an ATSEVA ANTAUS Yb-doped $1030$ nm fiber laser, at a repetition rate of $500$ kHz, are tightly focused on a $L = 10$ cm long borosilicate glass wafer (with refractive index $n_0 = 1.48$). The laser pulses weakly modify the material properties at the illuminated region, inducing a permanent refractive index contrast of $\Delta n\approx 10^{-4}-10^{-3}$~\cite{Guzm_n_Silva_2021}. Straight waveguides are created by translating the glass along the $z$ propagation coordinate by means of a motorized XYZ stage at a standard velocity of $20.0$ mm/s. This technique naturally produces vertically oriented waveguides due to the axial focusing of the laser.

\begin{figure}[h!]
  \center{\includegraphics[width=16.5cm]{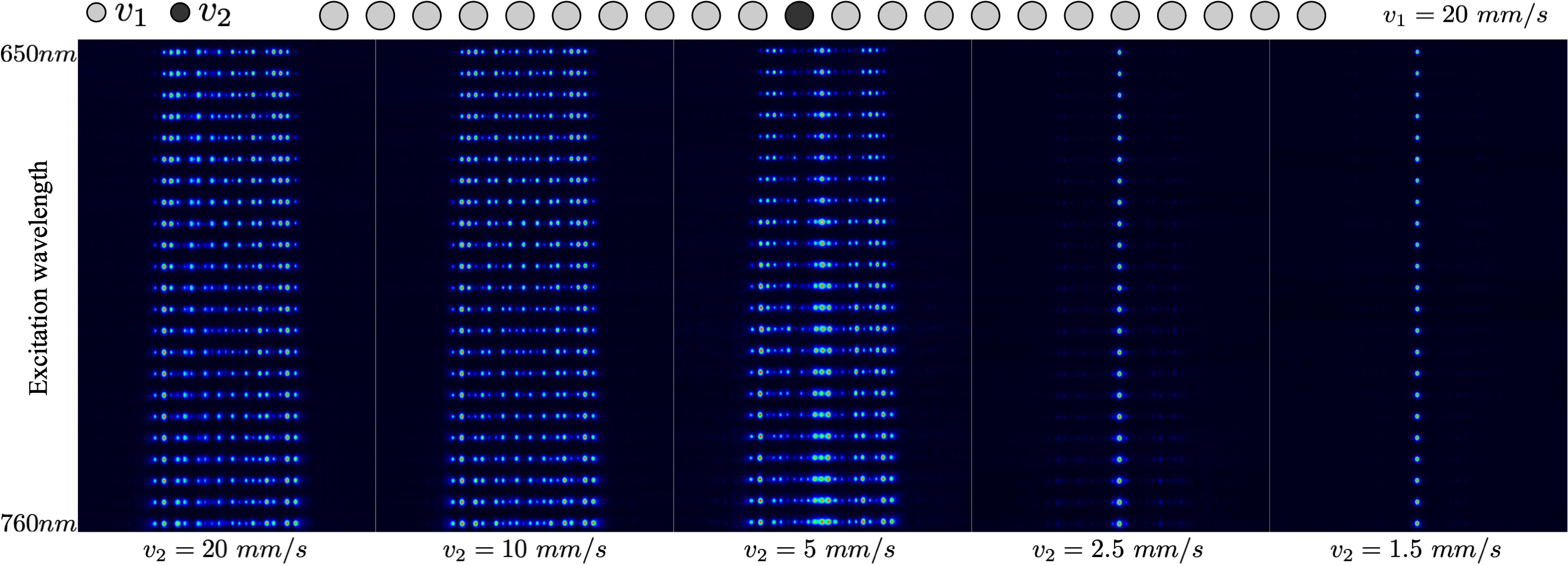}}
\caption{Top: Sketch of a 1D lattice with a single and central defect.  
Center: Compilation of experimental results for lattices with a decreasing (increasing) writing velocity (refractive index contrast) to the right, after the excitation of the central defect site, for different input wavelengths from 670 to 760 nm.}
\label{sm01}
\end{figure}

First of all, we study an on-site defect on a homogeneous 1D lattice as it is sketched in Fig.~\ref{sm01}-top. Lattice sites (light-gray) are fabricated with a nominal power of $30$ mW and a writing velocity $v_1=20$ mm/s. A defect waveguide (dark gray) is fabricated at the lattice center by changing the fabrication velocity. Figure~\ref{sm01} shows a compilation of results for different velocities $v_2$ (see below), after exciting the defect site at the input facet and using a different excitation wavelength, coming from a supercontinuum (SC) laser source. As commented in the previous section, we can modify the dynamical variable by using a longer glass or by increasing the coupling constants~\cite{C_ceres_Aravena_2023}.  We notice a very good transport (discrete diffraction) for a homogeneous lattice ($v_2=20$ mm/s) and only weak localization features at $v_2=5$ mm/s. Then, by decreasing further $v_2$ we notice a strong transition into a localized state, with a very weak background for $v_2=2.5$ mm/s and kind of perfect localization at $v_2=1.5$ mm/s.

\section{Disordered lattices: comparison between experiments and numerical simulations}

\begin{figure}[h!]
  \center{\includegraphics[width=12cm]{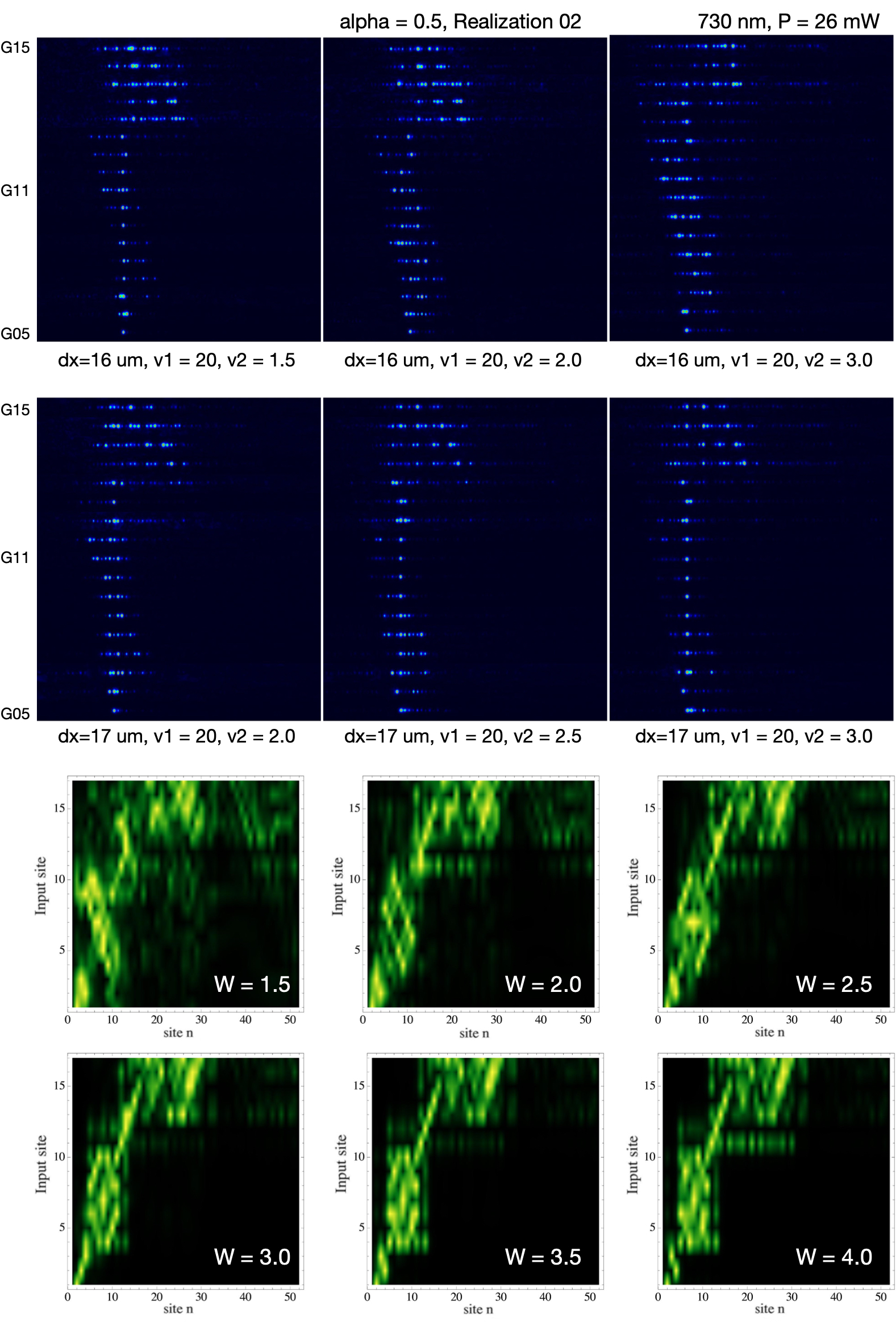}}
\caption{Top: Compilation of experimental results for a given realization (02) for $\alpha=0.5$ and for the indicated fabrication parameters, after the excitation of the indicated waveguides (G05--G15) for an input wavelength of $730$ nm. Bottom: Numerical compilation after integrating model (\ref{DLS}) and considering a coupling constant $\tau=1$, a propagation distance of $z=25$, and the indicated strength for the disorder $W$.} 
\label{sm02}
\end{figure}

After an initial experimental calibration, we proceed to optimize the lattice dimensions for the experiment. The goal is to have a maximum effective transport, which is the same as having a larger propagation distance. The larger glass lengths available in our laboratory are $L=10$ cm. Therefore, a good balance between coupling constants (directly determined by the waveguide distance~\cite{Szameit_2005}), with the on-site defects and the excitation wavelength is mandatory. Fig.~\ref{sm02}-top shows collected experimental results for $\alpha=0.5$ and a given realization (02), for a fixed input wavelength of $730$ nm, and a nominal fabrication power of $30$ mW. Every panel shows an output intensity profile after exciting the waveguides 5 to 15. In general, we observe similar results for $dx=16$ and $17\ \mu m$; therefore, we chose the smaller inter-site distance to enhance transport. Then, we notice that stronger defect waveguides (smaller $v_2$) produce stronger localization intensity patterns, although not many phenomenological differences are really observed. A direct comparison with numerical simulations (Fig.~\ref{sm02}-bottom) indicates that the experimental regimes are in good agreement with a weight $W\sim 3.0$ cm$^{-1}$ and for a propagation effective length of $z\approx 25$ cm, taking the coupling constant as $\tau=1$ cm$^{-1}$.

This calibration is crucial for this specific experiment because, in transport problems, many properties emerge more clearly at longer propagation distances and larger system sizes. However, available glasses define a first constraint: the maximum propagation length. Then, there is a way of effectively adjusting this parameter by increasing the coupling constants. As it was explained before, this can be done by reducing the intersite distances and/or by increasing the excitation wavelength. However, we also need to keep the first-nearest neighbor approximation valid in order to preserve the model under study. This immediately tells us that, at least experimentally, we need to fix the propagation length such that this approximation remains valid. Therefore, it makes no sense to look for longer glasses. In fact, considering the parameters already mentioned, we have been able to increase the real glass length from $10$ cm to an effective length of $25$ cm, which is indeed an enormous amplification. We could also use a longer excitation wavelength (e.g., $750$ nm) to increase the coupling and the effective distance; however, this would also decrease the strength of the defects, as well as increase next-nearest neighbours effects.

\section{Fabrication of Disorder Lattices}

After an exhaustive exploration of the available experimental parameters, we have established the following settings:
\begin{itemize}
    \item N=50
    \item $dx=16\ \mu$m
    \item $P_w=30$ mW
    \item $v_1=20$ mm/s
    \item $v_2=3$ mm/s
    \item $\lambda=730$ nm
    \item Horizontal polarization
\end{itemize}

\begin{figure}[h!]
  \center{\includegraphics[width=16cm]{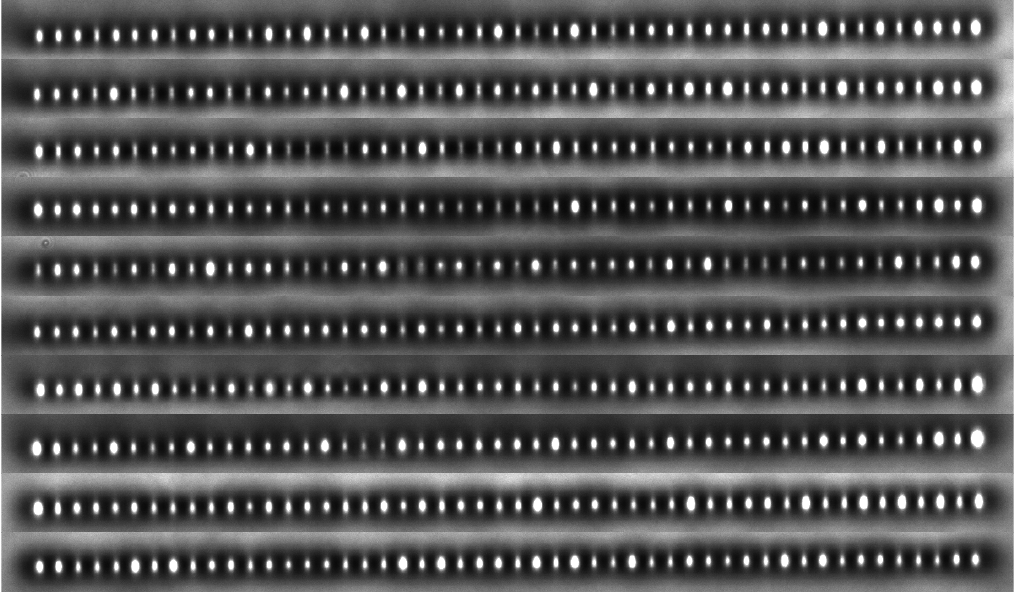}}
\caption{White light microscope images for some realizations of Anderson-like disordered ($\alpha=2.0$) samples.} 
\label{sm03}
\end{figure}

Fig.~\ref{sm03} shows a set of $\alpha=2.0$ disordered lattices, where the brighter and wider spots correspond to softer waveguides ($v_1$) while the thinner structures correspond to stronger waveguides ($v_2$).

\section{Direct measurement examples}

\begin{figure}[h!]
  \center{\includegraphics[width=16cm]{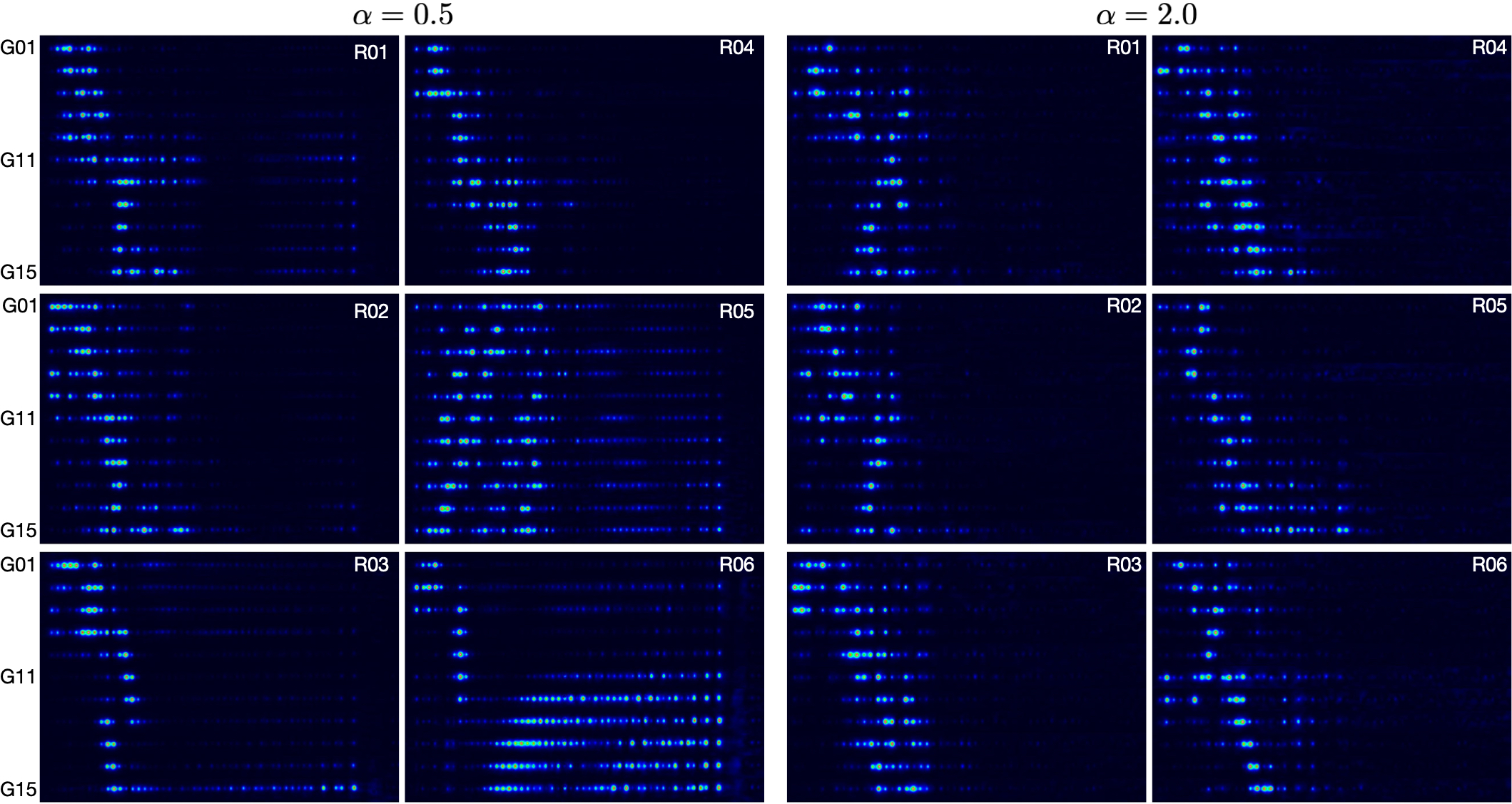}}
\caption{Compilation of experimental results for six realizations (R01 to R06), and for $\alpha=0.5$ and $2.0$ as indicated above. We show the excitation of 11 waveguides G05 to G15 ($y$-axis) with an excitation wavelength of $\lambda=730$ nm.}
\label{sm04}
\end{figure}

After a long calibration process, the final experiments consisted of exciting the fabricated lattices at 11 waveguides from 05 to 15. For each excitation, we obtained an intensity image on a CCD beam profiler. Figure~\ref{sm04} shows different panels with 11 excitations each, showing the study of every fabricated lattice. As every lattice is excited at 11 sites, and we fabricated a total of $38$ realizations for $\alpha=0.5$ and $2.0$, every $\alpha$ case has an ensemble of $418$ measurements. This is indeed a large number of measurements, where we tried to optimize as much as possible the use of every lattice to increase the statistics. 

\bibliography{Levy_Photonic_Bib}